\documentclass[twocolumn,showpacs,aps,prl,superscriptaddress]{revtex4}

\usepackage{graphicx}
\usepackage{dcolumn}
\usepackage{amsmath}
\usepackage{epsfig}

\input babarsym

\newcommand{\BABARPubYear}    {08}
\newcommand{\BABARPubNumber}  {029}

\newcommand{\SLACPubNumber} {13288}
\newcommand{\LANLNumber} {0807.1086}

\def\figurebox#1#2#3{%
    \def\arg{#3}%
    \ifx\arg\empty
    {\hfill\vbox{\hsize#2\hrule\hbox to #2{\vrule\hfill\vbox to #1{\hsize#2\vfill}\vrule}\hrule}\hfill}%
    \else
    {\hfill\epsfbox{#3}\hfill}%
    \fi}

\begin{document}

\preprint{\babar-PUB-\BABARPubYear/\BABARPubNumber} 
\preprint{SLAC-PUB-\SLACPubNumber} 

\begin{flushleft}
\babar-PUB-\BABARPubYear/\BABARPubNumber\\
SLAC-PUB-\SLACPubNumber\\
arXiv:\LANLNumber\ [hep-ex]\\[10mm]
\end{flushleft}

\title{
{\large \bf
   Observation of the bottomonium ground state in the decay
   $\Upsilon(3S) \to \gamma \, \eta_b$
 }
}

%
\author{B.~Aubert}
\author{M.~Bona}
\author{Y.~Karyotakis}
\author{J.~P.~Lees}
\author{V.~Poireau}
\author{E.~Prencipe}
\author{X.~Prudent}
\author{V.~Tisserand}
\affiliation{Laboratoire de Physique des Particules, IN2P3/CNRS et Universit\'e de Savoie, F-74941 Annecy-Le-Vieux, France }
\author{J.~Garra~Tico}
\author{E.~Grauges}
\affiliation{Universitat de Barcelona, Facultat de Fisica, Departament ECM, E-08028 Barcelona, Spain }
% ...
\collaboration{The \babar\ Collaboration}
\noaffiliation

\date{\today}

\begin{abstract}
We report the results of a search for the bottomonium ground state $\eta_b(1S)$
in the photon energy spectrum with a sample of $(109 \pm 1)$ million of $\Upsilon(3S)$ 
recorded at the $\Upsilon(3S)$ 
energy with the \babar\ detector at the PEP-II $B$ factory at SLAC.
We observe a peak in the photon energy spectrum at $E_\gamma = 921.2 ^{+2.1}_{-2.8} {\rm (stat)}\pm 2.4{\rm (syst)}$ MeV
with a significance of 10 standard deviations. We interpret the observed peak as being due to 
monochromatic photons from the radiative transition $\Upsilon(3S) \to \gamma \, \eta_b(1S)$.
This photon energy corresponds to an $\eta_b(1S)$ mass of $9388.9 ^{+3.1}_{-2.3} {\rm (stat)} \pm 2.7{\rm (syst)}$ MeV/$c^2$.
The hyperfine $\Upsilon(1S)$-$\eta_b(1S)$ mass splitting is $71.4 ^{+2.3}_{-3.1} {\rm (stat)} \pm 2.7{\rm (syst)}$ MeV/$c^2$.
The branching fraction for this radiative $\Upsilon(3S)$ decay is estimated to be
$(4.8 \pm 0.5{\rm (stat)}  \pm 1.2 {\rm (syst)}) \times 10^{-4}$.
\end{abstract}

\pacs{13.20.Gd, 14.40.Gx, 14.65.Fy}

\maketitle

Thirty years after the discovery of the narrow $\Upsilon(nS)$ resonances~\cite{ref:Ydiscovery},
no evidence has 
been reported for the spin-singlet pseudoscalar partners $\eta_b(nS)$ of these states.
Measurement of the hyperfine mass splittings between the triplet and singlet states in 
quarkonium systems is of key importance in understanding the role of spin-spin
interactions in quarkonium models and in testing QCD calculations~\cite{QWG-YR}.
Theoretical estimates of the mass splitting between the $1S$ 
singlet and triplet states vary from 36~MeV/$c^2$ to 100~MeV/$c^2$~\cite{ref:GodfreyRosner}.

In this letter, we report the observation of the radiative transition
$\Upsilon(3S) \to \gamma \, \eta_b(1S)$, where the $\eta_b(1S)$,
hereafter referred to as the $\eta_b$,
is the pseudoscalar partner of the triplet state $\Upsilon(1S)$,
and corresponds to the ground state of the bottomonium system.
Theoretical predictions of the decay branching fraction 
range from 1 to 20 $\times 10^{-4}$\cite{ref:GodfreyRosner}, 
where the unknown $\eta_b$ mass is a major source of the uncertainties.
The current limit from the CLEO III experiment,
${\cal B}[\Upsilon(3S)\to \gamma\, \eta_b] <4.3\times10^{-4}$ at 90\% confidence level, 
is based on 1.39 \invfb of $\Upsilon(3S)$ data~\cite{ref:cleo}.

The data sample used in this study was collected with the \babar\ detector~\cite{ref:babar} 
at the PEP-II asymmetric-energy $e^+e^-$ storage rings.  
It consists of 28.0 \invfb of integrated luminosity collected at a $e^+e^-$ center-of-mass (CM)
energy of 10.355~GeV, corresponding to the mass of the $\Upsilon(3S)$ resonance.
Additional samples of 2.4 \invfb and 43.9 \invfb were collected 30~MeV below
the $\Upsilon(3S)$ [below-$\Upsilon(3S)$] and 40~MeV below the $\Upsilon(4S)$ [below-$\Upsilon(4S)$] 
resonances, respectively and are used for background and calibration studies. 
The trajectories of charged particles are reconstructed
using a combination of five layers of double-sided silicon strip 
detectors and a 40-layer drift chamber, all operated inside the 1.5-T magnetic field
of a superconducting solenoid.
Photons are detected using a CsI(Tl) electromagnetic calorimeter 
(EMC), which is also inside the coil.
The energy resolution for photons varies from 2.9\% (at 600 MeV) to 2.5\% (at 1400 MeV).

 The signal for $\Upsilon(3S) \to \gamma \, \eta_b$ is extracted 
 from a fit to the inclusive photon energy spectrum in the CM frame.
 Any reference to photon energy hereafter will be in the CM frame, unless otherwise noted.

 The monochromatic photon from the decay appears as a peak
 on top of a smooth non-peaking background from
 continuum ($e^+e^- \to q\bar q$ with $q=u,d,s,c$) events and
 bottomonium decays.
 Two other processes produce peaks in the photon energy spectrum
 close to the signal region.
 Double radiative decays
 $\Upsilon(3S) \to \gamma \chi_{bJ}(2P); \chi_{bJ}(2P)\rightarrow\gamma\Upsilon(1S), \ J=0, 1, 2$, 
 produce a broad peak centered at 760 MeV due to photons from decays of the $\chi_{bJ}(2P)$ states.
 The peaks from the three $\chi_{bJ}(2P)$ transitions appear merged due to
 photon energy resolution and the Doppler broadening that arises from the motion of the $\chi_{bJ}(2P)$ 
in the CM frame.
 This $\chi_{bJ}(2P)$ photon peak is well separated from the signal region of interest (around $E_\gamma = 900$ MeV).
 We use the peak as a tool to verify the optimization of the selection criteria and to
 determine signal reconstruction efficiencies and the absolute photon energy scale.
 The other process leading to a peak near 860 MeV in the
 photon energy spectrum is the radiative production of the $\Upsilon(1S)$ via
 initial state radiation (ISR) $e^+e^- \to \gamma_{ISR} \, \Upsilon(1S)$.
 Knowledge of the magnitude and photon energy line shape of this background is
 crucial in extracting the $\eta_b$ signal.

We employ a simple set of selection criteria to suppress the backgrounds
while retaining a high signal efficiency.
Decays of the $\eta_{b}$ via two gluons, expected to be a large component of its decay modes, have 
high track multiplicity. 
Hadronic events are selected by requiring four or more charged tracks in the event and 
that the ratio of the second to zeroth Fox-Wolfram moments~\cite{ref:fox} be less than 0.98. 

Photon candidates are required to be isolated from all charged tracks.
To ensure that their shapes are consistent with an electromagnetic shower,
the lateral moments~\cite{ref:LAT} are required to be less than 0.55.
The signal photon candidate is required to lie in the central angular region of the EMC,
$-0.762<\cos(\theta_{\gamma, LAB})<0.890$, 
where $\theta_{\gamma, LAB}$ is the angle between the photon and the beam axis in the 
laboratory frame. This requirement ensures high reconstruction efficiency and good energy resolution,
and reduces the contributions of ISR photons from $e^+ e^- \rightarrow \gamma_{ISR} \Upsilon(1S)$ 
events.

Due to the fact that there is no preferred direction in the decay of the spin-zero $\eta_b$, 
the correlation of the direction of the photon momentum in the CM frame with the 
thrust axis~\cite{ref:brandt} of the $\eta_b$ is small. 
In contrast, there is a strong correlation between the photon direction
and thrust axis in continuum events.
The thrust axis is computed with all charged tracks
and neutral calorimeter clusters in the event, with the exception of the
signal photon candidate.
We require $|\cos{\theta_T}|< 0.7$ to reduce continuum background,
where $\theta_T$ is the angle between the thrust axis and the signal photon
candidate in the CM frame.

Photons from \piz decays are one of the main sources of background.
A signal photon candidate is rejected if it combines with another photon
in the event to form a \piz candidate within 15 MeV/$c^2$ of the nominal \piz mass.
To maintain high signal efficiency,
we require the second photon of the \piz candidate
to have an energy in the laboratory frame greater than $50~\mev$.

The above mentioned selection criteria were chosen by optimizing
the $S/\sqrt(B)$ ratio between the expected signal yield ($S$) and
the background ($B$). The signal sample in the optimization
is provided by a detailed Monte Carlo (MC) simulation~\cite{ref:geant}. 
Since no reliable event generators exist to simulate the background photon distribution, 
especially from bottomonium decays, a small fraction (9\%) of the $\Upsilon(3S)$ data is used 
in the optimization to model the background in the region $0.85<E_\gamma<0.95$\gev. 
To avoid potential bias, these data are not used in the final fit
of the photon energy spectrum. 
This optimization procedure, when applied to the $\chi_{bJ}(2P)$ yield in data 
in place of the simulated signal, yields the same optimal selection criteria.
The final reconstruction efficiency evaluated from the simulated signal MC events is 37\%. 

The remaining $\Upsilon(3S)$ data used for the analysis has an integrated luminosity of 25.6 \invfb,
which corresponds to $(109 \pm 1)$ million $\Upsilon(3S)$ events.

To extract the $\eta_b$ signal, we perform a binned maximum likelihood (ML) fit of the $E_\gamma$ spectrum 
with $0.5<E_\gamma<1.1$\gev with four components: 
non-peaking background,
$\chi_{bJ}(2P)\rightarrow \gamma \, \Upsilon(1S)$, $\gamma_{ISR} \Upsilon(1S)$, and the $\eta_b$ signal.  

The non-peaking background
is parametrized by the following probability density function (PDF), 
$f(E_{\gamma}) = A \left( C + {\rm exp}[-\alpha E_\gamma - \beta E^2_\gamma] \right)$.

The form of the $\chi_{bJ}(2P)$ PDF is complicated 
by the presence of Doppler broadening.
Crystal Ball (CB) functions~\cite{ref:CB} are used as phenomenological PDFs for the three 
$\chi_{bJ}(2P)\rightarrow \gamma \Upsilon(1S)$ shapes.
The CB function is a Gaussian modified to have an extended, power-law tail
on the low (left) side.
The relative rates and peak positions of the $\chi_{bJ}(2P)$ components are fixed to their 
world-averaged (PDG) values~\cite{ref:PDG}.
 The parameters describing the low-side tail of the CB function
 are common to all three of the $\chi_{bJ}(2P)$ peaks.
The $\chi_{bJ}(2P)$ PDF parameters are determined by fitting the photon
energy spectrum,
with the signal region (840 to 960 MeV) excluded,
after subtraction of the non-peaking  background.
All of the $\chi_{bJ}(2P)$ PDF  parameters from this fit,
with the exception of the overall normalization, are fixed in the ultimate
fit to the full photon energy spectrum.

\begin{figure}[htb]
\begin{center}
\includegraphics[scale=0.40]{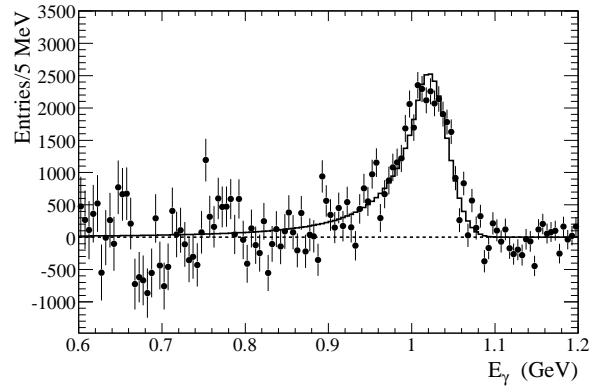}
\end{center}
\caption{\label{fig:off4S} 
Inclusive photon energy spectrum in the below-$\Upsilon(4S)$ data, with the non-peaking 
background subtracted.The peak at 1.03 GeV is from the ISR process 
$e^+e^- \to \gamma_{ISR} \, \Upsilon(1S)$. The superimposed histogram corresponds to a 
fit with a CB function.}
\end{figure}

The PDF of the peaking background from ISR $\Upsilon(1S)$ production 
is parametrized as a CB function form whose parameters are determined from simulated events.
To estimate the rate of this continuum component in $\Upsilon(3S)$ data,
we use the below-$\Upsilon(3S)$ and below-$\Upsilon(4S)$ data.
Figure 1 shows the $E_\gamma$ distribution in the below-$\Upsilon(4S)$ data, after 
applying the selection criteria and subtracting the non-peaking background. The fit 
with a CB function yields $35800 \pm 1600$ events. Extrapolating the cross section 
to the $\Upsilon(3S)$ energy and correcting for the luminosity ratio and the small difference 
in detection efficiency, the ISR photon background contribution in the 
final analysis is estimated to be $25200 \pm 1700$ events. The error includes 
systematic uncertainties. This is consistent with and more precise 
than the rate estimated using the below-$\Upsilon(3S)$ data.

\begin{figure}
\begin{center}
\includegraphics[scale=0.45]{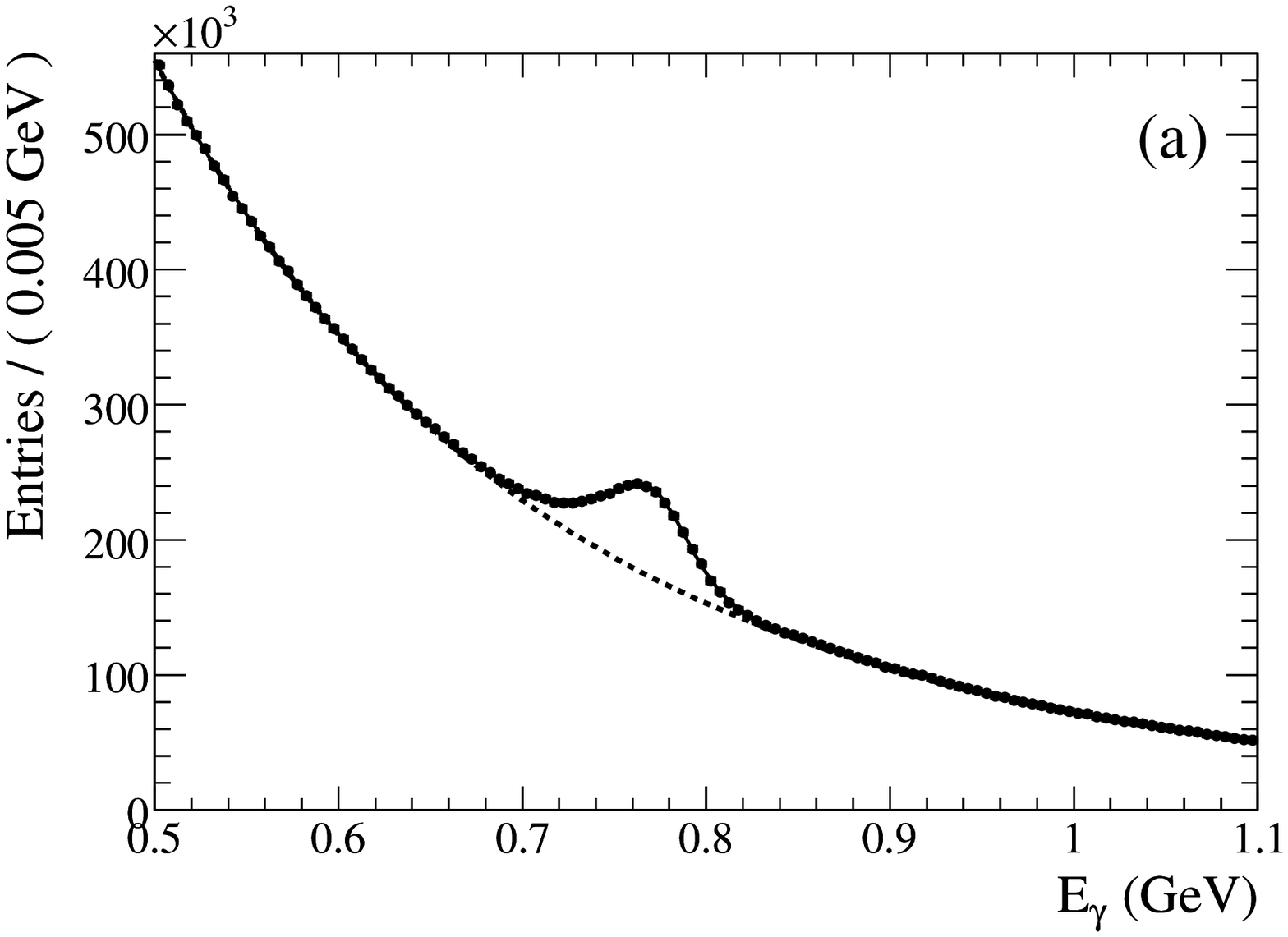}
\includegraphics[scale=0.45]{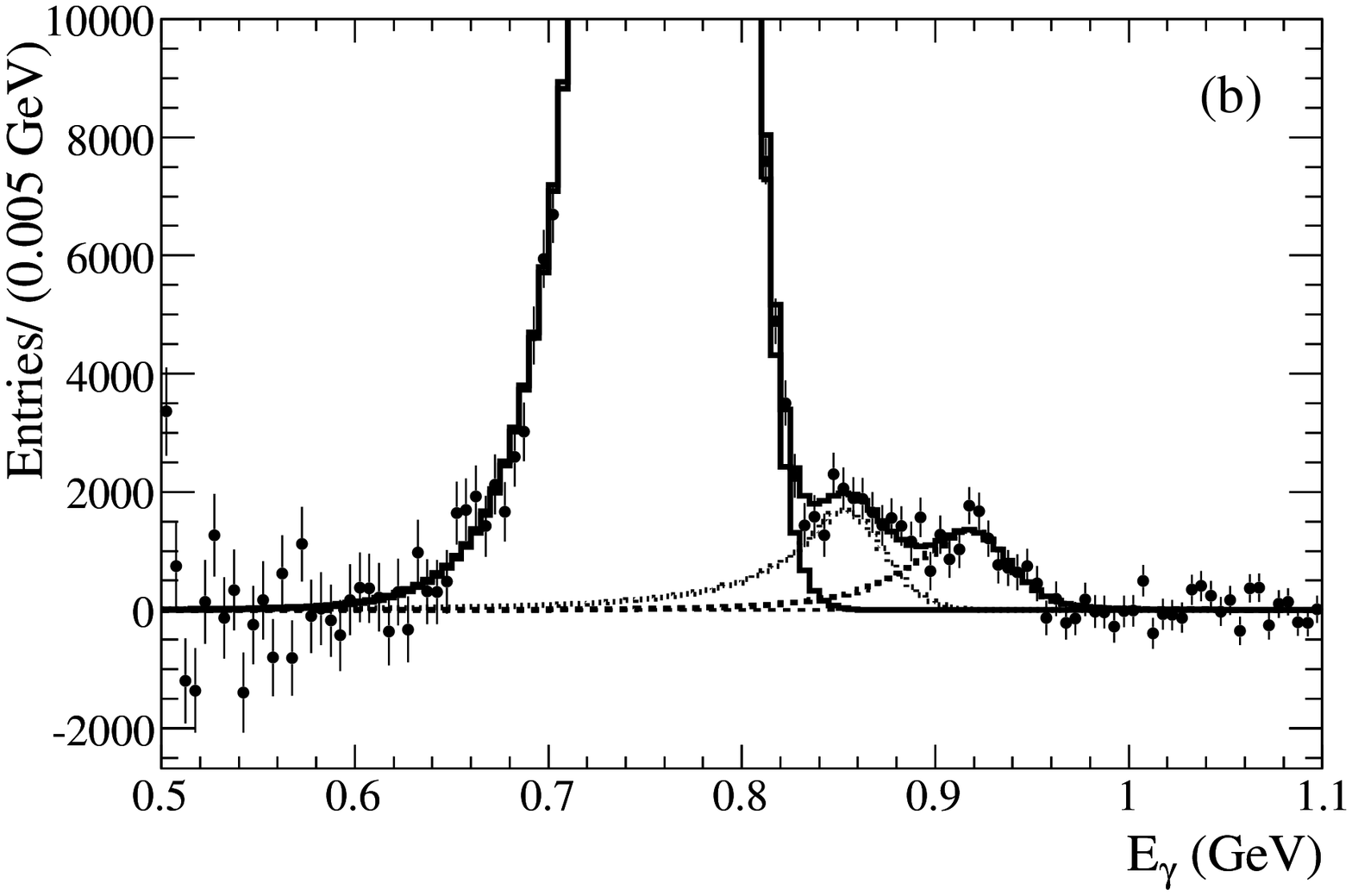}
\includegraphics[scale=0.45]{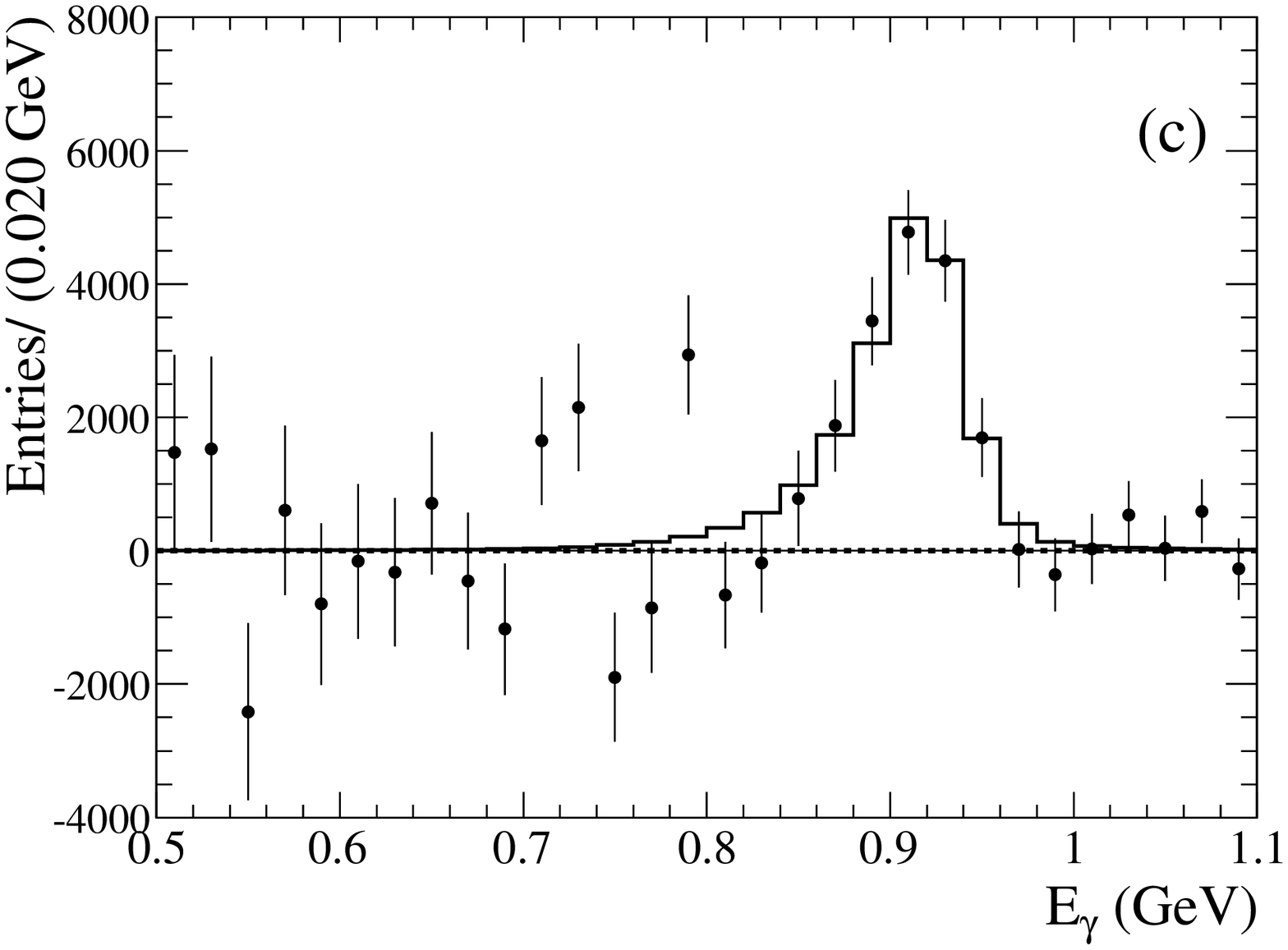}
\end{center}
\caption{\label{fig:backgroundsubtracted} 
(a) Inclusive photon spectrum in the region $0.50<E_\gamma<1.1$\gev.
    The component PDFs determined from the fit are overlaid on the data points.
    A prominent $\chi_{bJ}(2P)$ peak is clearly seen. The dashed line corresponds 
to the non-peaking background component. 
(b) Inclusive photon spectrum after subtracting the non-peaking background,
with PDFs for $\chi_{bJ}(2P)$ peak (solid), ISR $\Upsilon(1S)$ (dot), $\eta_b$ signal (dash) and 
the sum of all three (solid). 
(c) Inclusive photon spectrum after subtracting all components except the 
$\eta_b$ signal. The CB function shape describes the data points well.
}
\end{figure}

In the final fit of the whole $E_\gamma$ distribution to extract the $\eta_b$ signal,
all parameters of the $\chi_{bJ}(2P)$ peak and the ISR $\Upsilon(1S)$ PDFs are
fixed to the values from the fits described above.

The $\eta_b$ signal PDF is a non-relativistic Breit-Wigner convolved with
a CB function to account for the experimental $E_\gamma$ resolution.
The CB parameters are determined from signal MC with the
$\eta_b$ width set to zero.
Since the width of the $\eta_b$ is not known, we have chosen a nominal value of 10 MeV/$c^2$
for the width. Theoretical predictions based on the expected ratio of 
the two-photon and two-gluon widths range from 4 to 20 MeV~\cite{ref:widththeory}.
The free parameters in the fit are the $\eta_b$ peak position and signal yield,
the $\chi_{bJ}(2P)$ yield, and all of the non-peaking background PDF parameters.

Figure 2(a) shows the photon energy spectrum and the fit result. The non-peaking background 
is dominant with only the prominent $\chi_{bJ}(2P)$ peak visible. In Figure 2(b) we 
show the detail of the signal region, after subtracting the non-peaking background. 
The line shapes of the three peaking components, $\chi_{bJ}(2P)$, ISR $\Upsilon(1S)$, and the 
$\eta_b$ signal are clearly visible. The $\chi^2$ per degree of freedom from the fit is 147/113 = 1.3. 
Finally Figure 2(c) shows the data points with all components except the $\eta_b$ signal  
subtracted, overlaid with the $\eta_b$ signal PDF. 
The fitted $\eta_b$ signal yield is $19200 \pm 2000 \pm 2100$ events, 
where the first error is statistical and the second systematic.
A total systematic uncertainty of 11\% is estimated by varying
the Breit-Wigner width  in the $\eta_b$ PDF to 5, 15, and 20 MeV,
setting the ISR $\Upsilon(1S)$ component to $\pm$1 $\sigma$ of the nominal rate,
and varying the PDF parameters fixed in the fit by $\pm$1 $\sigma$. 
The largest contribution (10\%) is from the $\eta_b$ width variation.

The $\eta_b$ signal significance is estimated using the ratio  log$ (L_{\rm max} / L_0)$, 
where $L_{\rm max}$ and $L_0$ are the likelihood values obtained from the nominal fit and from 
a fit with the $\eta_b$ PDF removed, respectively. 
Fits have been performed where the parameters entering the systematic uncertainties 
have been varied within their errors. Data have then been fitted with all parameters 
simultaneously moved by one standard deviation in the direction of lower significance. 
This conservative approach yields a signal significance greater than 10 standard deviations.

As a cross check, we also perform a fit where the yield of the ISR $\Upsilon(1S)$ component is 
left free, and we 
obtain $24800 \pm 2300$ events for this component. This is consistent with the estimate 
using the below-$\Upsilon(4S)$ data and provides an important validation of the 
$\chi_{bJ}(2P)$ line shape parameterization. The yield and peak position of the $\eta_b$ 
signal from this fit are unchanged.

The $E_\gamma$ signal peak value from the fit is $917.4 ^{+2.1}_{-2.8}$ MeV.
We apply a photon energy calibration shift of $3.8 \pm 2.0 $ MeV, 
obtained by comparing the fitted position of the $\chi_{bJ}(2P)$ peak to 
the known PDG value. 
After including an additional systematic uncertainty of 1.3 MeV from 
the fit variations described above, we obtain a value of
$E_\gamma = 921.2 ^{+2.1}_{-2.8} \pm 2.4$ MeV for the $\eta_b$ signal. 

The $\eta_b$ mass derived from the $E_\gamma$ signal is
$M(\eta_b) = 9388.9 ^{+3.1}_{-2.3} \pm 2.7$ MeV/$c^2$.
Using the PDG value of $9460.3 \pm 0.3$ MeV/$c^2$ for the $\Upsilon(1S)$ mass,
we determine the $\Upsilon(1S)$-$\eta_b$ mass splitting to be
$71.4 ^{+2.3}_{-3.1} \pm 2.7$ MeV/$c^2$.

The value we measure for the splitting is larger than most predictions
based on potential models [2], but reasonably in agreement with predictions 
from lattice calculations~\cite{Gray:2005ur}. 
The mass splitting between the \OneS\ and the $\eta_b(1S)$ is a key
ingredient in many theoretical calculations. The precision of our
measurement will allow, among others, a more precise determination of
the lattice spacing~\cite{Gray:2005ur}  and new precision determinations
of $\alpha_s$~\cite{Kniehl:2003ap}.

We estimate the branching fraction by correcting the signal yield with the reconstruction
efficiency ($\epsilon$) from simulated signal MC events, and then dividing it by 
the number of $\Upsilon(3S)$ events in the data sample.
The branching fraction of the decay $\Upsilon(3S) \to \gamma \, \eta_b$ is found 
to be $(4.8 \pm 0.5 \pm 1.2) \times 10^{-4}$, where the first uncertainty is statistical
and the second systematic.
The systematic uncertainty of 25\% comes from uncertainties in 
the signal yield (11\%) and $\epsilon$ (22\%).
The latter is obtained by comparing the yield of $\chi_{bJ}(2P)$ in data to 
the number of expected events, which is calculated from the known branching 
fractions~\cite{ref:PDG}, the number of $\Upsilon(3S)$ events, and 
MC reconstruction efficiency of $\chi_{bJ}(2P)$.
They show a 13\% discrepancy, but are consistent within the errors. 
We assign the full difference to the systematic uncertainty.
A total uncertainty in $\epsilon$  is obtained, after adding the uncertainties 
in the $\chi_{bJ}(2P)$ branching fractions (18\%).

In conclusion, we have observed, with a significance of 10 standard
deviations, the radiative decay of the $\Upsilon(3S)$ to a narrow state lying
slightly below the $\Upsilon(1S)$. The most likely interpretation of the signal 
peak is the $\Upsilon(3S)$ transition to the bottomonium ground state, although 
other hypotheses, such as a radiative transition to a light Higgs boson, are not 
excluded. Under the bottomonium interpretation, this is the first evidence for 
the $\eta_b$ bottomonium state, the  pseudoscalar partner of the $\Upsilon(1S)$. 
The mass of the $\eta_b$ is $9388.9 ^{+3.1}_{-2.3} \pm 2.7$ MeV/$c^2$, which 
corresponds to a mass splitting between the $\Upsilon(1S)$ and the $\eta_b$ of 
$71.4 ^{+2.3}_{-3.1} \pm 2.7$ MeV/$c^2$. The estimated branching fraction of the decay 
$\Upsilon(3S) \to \gamma \, \eta_b$ is found to be $(4.8 \pm 0.5 \pm 1.2) \times 10^{-4}$.

We are grateful for the excellent luminosity and machine conditions
provided by our \pep2\ colleagues, 
and for the substantial dedicated effort from
the computing organizations that support \babar. We thank Bob McElrath 
and Michael Peskin for helpful discussions. 
The collaborating institutions wish to thank 
SLAC for its support and kind hospitality. 
This work is supported by
DOE
and NSF (USA),
NSERC (Canada),
CEA and
CNRS-IN2P3
(France),
BMBF and DFG
(Germany),
INFN (Italy),
FOM (The Netherlands),
NFR (Norway),
MES (Russia),
MEC (Spain), and
STFC (United Kingdom). 
Individuals have received support from the
Marie Curie EIF (European Union) and
the A.~P.~Sloan Foundation.

\end{document}